\documentclass[letterpaper,twocolumn,showpacs,prl,aps]{revtex4-1}

\usepackage{amsmath}  
\usepackage{amsfonts} 
\usepackage{graphicx} 
\usepackage{amssymb}

\begin{document}

\title{Nonlinear plasma waves in an electron gas}

\author{Eugene B. Kolomeisky}

\affiliation
{Department of Physics, University of Virginia, P. O. Box 400714, Charlottesville, Virginia 22904-4714, USA}

\date{\today}

\begin{abstract}
The nature of traveling wave solutions to equations of hydrodynamics of a three-dimensional electron gas with parabolic dispersion law depends on whether the motion is subsonic or supersonic.  Solitons representing localized depressions of the electrostatic potential and electron density are predicted to exist in the subsonic regime:  at rest the solitons are dark while in motion they are grey.  Two types of periodic waves are found in the supersonic regime: (i) smooth waves whose small amplitude limit is described by harmonic theory, and (ii) waves with sharp troughs and smooth crests of the potential with the electrons accumulating in the troughs.                 
\end{abstract}


\maketitle

An interacting electron gas in the presence of a uniform positively charged background is one of the paradigms that has framed our understanding of the physics of metals and doped semiconductors \cite{Bohm_Pines,Pines_Nozieres,Mahan}.  It provides a reasonable approximation to real materials, correctly capturing the ground-state properties,  screening, and the excitation spectrum.  Long-wavelength properties of the electron gas can be derived from a macroscopic theory \cite{Feynman,Jackson,Fetter} that dates back to Bloch's hydrodynamic generalization of the Thomas-Fermi theory of a neutral atom \cite{Bloch}.  In this approach the electrons are treated as charged ideal liquid characterized by the local position- and time-dependent number density $n(\textbf{r},t)$ and velocity $\textbf{v}(\textbf{r},t)$ fields, which are related by the continuity equation
\begin{equation}
\label{continuity}
\frac{\partial n}{\partial t}+\nabla \cdot(n\textbf{v})=0
\end{equation}
The equation of motion of the liquid is given by the Euler equation of hydrodynamics 
\begin{equation}
\label{2nd_law}
m\left [\frac{\partial \textbf{v}}{\partial t} +(\textbf{v}\cdot \nabla)\textbf{v}\right ]=-\nabla \left [\zeta(n) - e \varphi\right ]
\end{equation}
where $m$ is the effective electron mass and $\zeta(n)$ is the chemical potential of the electrons in the absence of the electrostatic potential $\varphi$.  

Since the electrons are significantly slower than light, the effects of retardation are neglected, i.e. the potential $\varphi$ is determined by the density $n$ via the Poisson equation
\begin{equation}
\label{Poisson}
\nabla^{2}\varphi=4\pi e(n-n_{0})
\end{equation}
where $n_{0}$ is the number density of the neutralizing background \cite{note}.  Linearizing Eqs.(\ref{continuity})-(\ref{Poisson}) about the state of equilibrium, $n=n_{0}$, $\textbf{v}=0$, $\varphi=0$, predicts that small oscillations of the electron liquid, the plasma oscillations, are harmonic with the dispersion law $\Omega(\textbf{q})$ given by \cite{Feynman,Jackson,Fetter}
\begin{eqnarray}
\label{spectrum}
\Omega^{2}(\textbf{q})&=&\omega_{0}^{2}+ s^{2}q^{2}\equiv \omega_{0}^{2}(1+d^{2}q^{2}),\nonumber\\\omega_{0}^{2}&=&\frac{4\pi n_{0}e^{2}}{m},~~s^{2}=\frac{n_{0}}{m}\left (\frac{\partial \zeta}{\partial n}\right )_{0},~~d=\frac{s}{\omega_{0}}
\end{eqnarray}
where $\Omega$ is the frequency of an excitation of the wave vector $\textbf{q}$, $\omega_{0}$ is the classical plasma frequency \cite{Tonks}, $s$ is the speed of sound in the neutral $e=0$ limit \cite{LL6}, and $d$ is the Debye screening length of the electron gas.  Pertinent information about the electron system such as interactions, thermal and quantum-mechanical effects is accumulated in the equation of state $\zeta(n)$.  Chemical potential of the electron gas $\zeta$ is a monotonically increasing function of its density $n$, and a range of cases is captured by the polytropic law
\begin{equation}
\label{equation_of_state}
\zeta(n)=\frac{ms^{2}}{\gamma-1}\left (\frac{n}{n_{0}}\right )^{\gamma-1}
\end{equation}  
where $\gamma$ is the polytropic index.  For example, $\gamma=5/3$ describes non-interacting Fermi gas \cite{LL5} in the low-frequency limit;  in the opposite high-frequency case one has to employ $\gamma=3$ \cite{SP,Jackson, Fetter}.  The $\gamma=3$ equation of state will be often used below to illustrate statements of general nature. 

It has been proposed \cite{Manfredi,Shukla} to explicitly account for quantum-mechanical effects by amending the right-hand side of the Euler equation (\ref{2nd_law}) according to $\zeta\rightarrow \zeta+Q$ where $Q=-(\hbar^{2}/2m)\nabla^{2}(\sqrt{n})/\sqrt{n}$ is the quantum potential that appears in the Madelung transformation of the Schr\"odinger equation \cite{Madelung} and Bohm's interpretation of quantum mechanics \cite{Bohm}.  In such modification called quantum hydrodynamics (QHD) model \cite{Manfredi,Shukla} the strength of quantum-mechanical effects is quantified by the $(Q/\zeta)_{0}$ ratio which can be estimated as
\begin{equation}
\label{quantum}
\mathcal{A}=\left (\frac{Q}{\zeta}\right )_{0}\simeq\frac{\hbar^{2}}{md^{2}\zeta(n_{0})}\simeq\left[\frac{\hbar \omega_{0}}{\zeta(n_{0})}\right ]^{2}
\end{equation}
The QHD model may be viewed as Bloch's hydrodynamics based on modified Thomas-Fermi theory with von Weizs\"acker's quantum correction included \cite{von}.  Such an improvement is known to be uncontrolled because it overlooks the exchange interaction that has the same order of magnitude effect as the quantum correction;  the latter is also known to be too large by a factor of $9$ \cite{correct}.  For these reasons the QHD model is not pursued below.      

While the harmonic approximation is very useful and simple to deal with, it is insufficient whenever the electron gas is subject to strong perturbations.  In this paper we classify traveling wave solutions to the system of Eqs.(\ref{continuity})-(\ref{Poisson}) without  assuming that the wave amplitude is small.  For the special case of an infinite compressibility ($\zeta=const$) gas this problem has been solved in a pioneering paper of Akhiezer and Lyubarskii (AL) \cite{AL,SZ}.  Here we show that restoring finite compressibility of the electron gas qualitatively modifies the AL results.  

We focus on the case of a one-dimensional motion along the $x$ axis, $\textbf{v}=(v,0,0)$, and seek solutions for the density $n$, velocity $v$ and electrostatic potential $\varphi$ that depend only on $\xi=x-ut$, i.e. having a form of a wave propagating in the positive $x$ direction with velocity $u$.  Then Eqs.(\ref{continuity})-(\ref{Poisson}) transform into 
\begin{equation}
\label{1d_continuity}
-un'+(nv)'=0
\end{equation}
\begin{equation}
\label{1d_Euler}
m (-uv'+vv')=-(\zeta-e\varphi)'
\end{equation}
\begin{equation}
\label{1d_Poisson}
\varphi''=4\pi e(n-n_{0})
\end{equation}
where the prime stands for the derivative with respect to $\xi$.  We limit ourselves to the waves containing a point where $v=0$ and thus $\varphi=0$ and $n=n_{0}$.  Using these as boundary conditions in Eqs.(\ref{1d_continuity}) and (\ref{1d_Euler}) and integrating one finds 
\begin{equation}
\label{1d_continuity_integrated}
n=n_{0}\frac{u}{u-v}
\end{equation}
\begin{equation}
\label{1d_Euler_integrated}
\frac{mu^{2}}{2}+\zeta(n_{0})=\frac{m(u-v)^{2}}{2}+\zeta(n)-e\varphi
\end{equation}
First of these implies that the electrons are slower than the wave, $v<u$, while the second generalizes also known result \cite{AL,SZ} to the case of $\zeta\neq const$.  In the reference frame of the wave the electrons flow over static potential energy landscape $-e\varphi$ in the negative $x$ direction, and  Eq.(\ref{1d_Euler_integrated}) is a statement of conservation of energy accounting for electron's "internal" energy $\zeta(n)$.  Eqs.(\ref{1d_continuity_integrated}) and (\ref{1d_Euler_integrated}) can be combined into an expression
\begin{equation}
\label{combined}
\frac{mu^{2}}{2}\left (1- \frac{n_{0}^{2}}{n^{2}}\right )=\zeta(n)-\zeta(n_{0})-e\varphi
\end{equation}
which can be solved to infer potential dependence of the electron density $n(\varphi)$ that appears in the right-hand side of the Poisson equation (\ref{1d_Poisson}).  If $\varphi$ is viewed as a position of a fictitious classical particle of unit mass, $\xi$ as a time, and $4\pi e[n(\varphi)-n_{0}]$ as an exerted force, then Eq.(\ref{1d_Poisson}) parallels Newton's second law of motion for the particle.  The first integral of (\ref{1d_Poisson}), the "energy" integral, then has the form
\begin{equation}
\label{energy_integral}
\frac{\varphi'^{2}}{2}+\mathcal{S}(\varphi)=E,~~~\mathcal{S}(\varphi)=4\pi e \int_{0}^{\varphi}[n_{0}-n(\varphi)]d\varphi
\end{equation}  
where the integration constant $E$ is the "energy" while $\mathcal{S}(\varphi)$ is the "potential energy"; hereafter without the loss of generality we choose $\mathcal{S}(0)=0$.  

Given $\mathcal{S(\varphi)}$, traveling wave solutions of Eqs.(1)-(3) can be classified according to possible motions of classical particle in the field of potential energy $\mathcal{S}(\varphi)$ for different values of the energy $E$.  This idea has been pioneered by Sagdeev in plasma physics \cite{SP};  hereafter the function $\mathcal{S}(\varphi)$ is called the Sagdeev potential (SP) \cite{tools}.  Below we will be interested in finite motions in the SP corresponding to finite amplitude waves in the electron gas.  

If $\varphi=0$ is a minimum of $\mathcal{S}(\varphi)$,  it is required that $E\geqslant0$ for the motion to be finite.    This scenario is realized in the $\zeta=const$ gas \cite{AL,SZ}.  Indeed solving Eq.(\ref{combined}) relative to the density and substituting the outcome into the Poisson equation (\ref{1d_Poisson}) we find
\begin{equation}
\label{AL_Poisson}
\psi''=(1+2\psi)^{-1/2}-1, ~~~\psi\geqslant -\frac{1}{2}
\end{equation}
where $\psi=e\varphi/mu^{2}$ is dimensionless potential and we measure length in units of $\lambda=u/\omega_{0}$ which sets the spatial scale of the waves.  The first integral of (\ref{AL_Poisson}) is given by
\begin{equation}
\label{AL_Sagdeev}
\frac{\psi'^{2}}{2}+\mathcal{S}(\psi)=E,~~~\mathcal{S}(\psi)= 1+\psi-\sqrt{1+2\psi}
\end{equation} 
The SP $\mathcal{S}(\psi)$ has a minimum at $\psi=0$ and finite motions can be inferred from the phase portrait of the system (\ref{AL_Sagdeev}) shown in Figure 1:
\begin{figure}
\includegraphics[width=1.0\columnwidth, keepaspectratio]{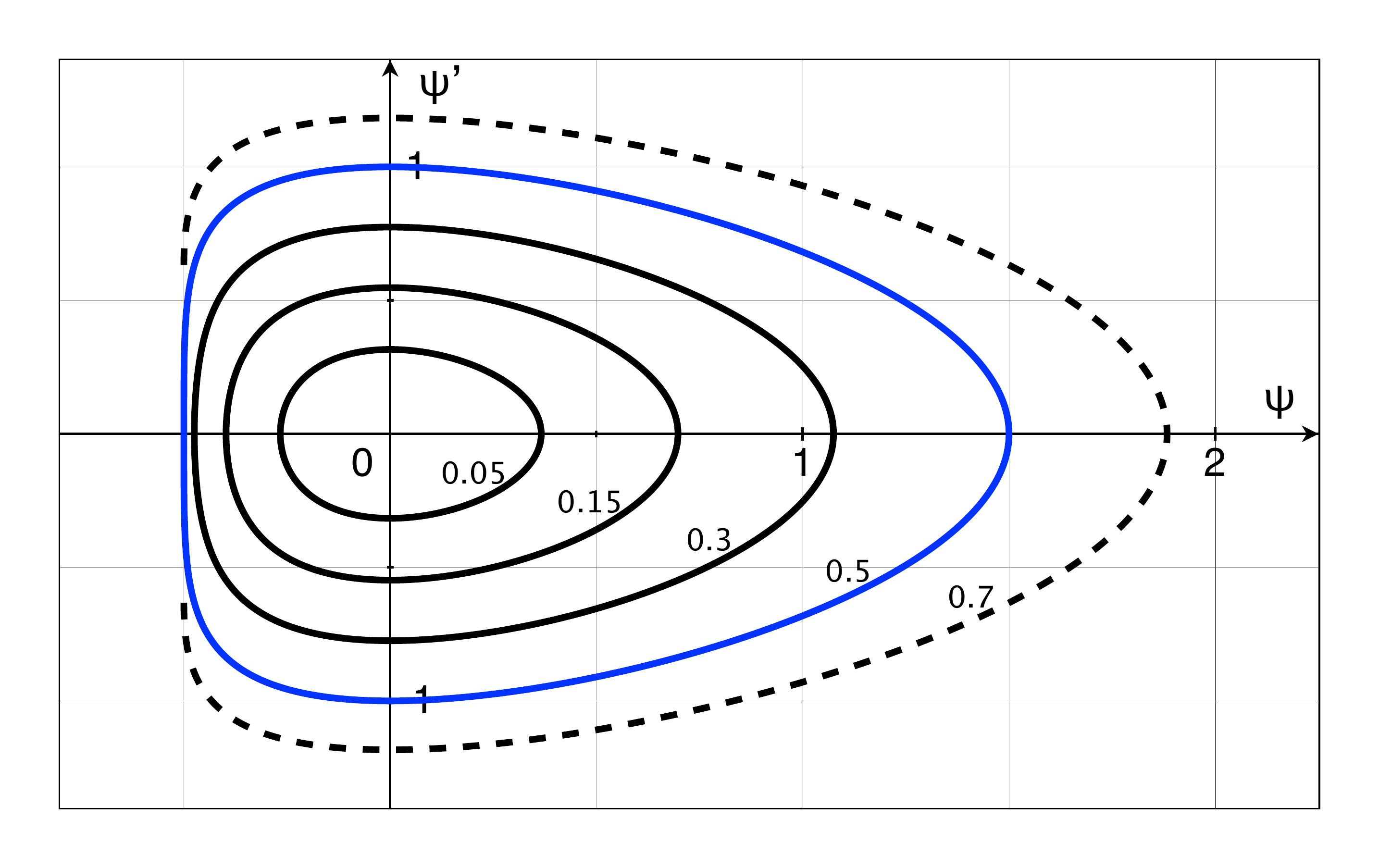} 
\caption{(Color online) Phase portrait of plasma oscillations of an electron gas of infinite compressibility.  Numbers next to phase trajectories are values of the integration constant $E$ in Eq.(\ref{AL_Sagdeev}).}
\end{figure}

While very small $E$ phase trajectories are close to elliptical (i.e. corresponding oscillations are harmonic), anharmonic effects are readily noticeable at sufficiently small $E$.  The phase trajectory corresponding to $E=0.5$ shown in blue is marginal because this is where the dimensionless potential $\psi$ takes on its most negative value of $-1/2$ for the first time.  At this point the velocity $v$ reaches the velocity of the wave $u$ and the electron density (\ref{1d_continuity_integrated}) diverges.  This is the effect of wave breaking  \cite{SZ,SP}.  Upon further increase of $E$ the phase trajectory (see dashed $E=0.7$ curve in Figure 1) develops discontinuity in $\psi'$ at $\psi=-1/2$:  this is where fictitious particle moving in the SP (\ref{AL_Sagdeev}) begins experiencing hard wall reflection.  As a result, the potential profile develops sharp minima at $\psi=-1/2$;  the wave is characterized by sharp troughs and smooth crests.   While the wave breaking eliminates traveling wave solutions to Eqs.(\ref{continuity})-(\ref{Poisson}) for $E\ge0.5$ \cite{SZ}, the effect itself is an artifact of the $\zeta=const$ model.  In a realistic situation unlimited electron accumulation in the troughs of the potential is halted by finite compressibility of the electron gas and counterparts to the $E\geqslant 0.5$ oscillations become accessible as is shown below.

If $\varphi=0$ is a maximum of the SP, then it is required that $E\leqslant 0$ for the motion to be finite.  This scenario is realized in a generic electron gas of finite compressibility in the static $u=0$ limit. Then $v=0$ and Eqs.(\ref{1d_Euler_integrated}) and (\ref{combined}) reduce to the condition of mechanical equilibrium $\zeta(n_{0})=\zeta(n)-e\varphi$.  Employing the $\gamma=3$ equation of state (\ref{equation_of_state}), solving relative to the density and substituting the outcome into the Poisson equation (\ref{1d_Poisson}) we find
\begin{equation}
\label{static_Poisson}
\phi''=\sqrt{1+2\phi}-1, ~~~\phi\geqslant-\frac{1}{2}
\end{equation}
where $\phi=e\varphi/ms^{2}$ is dimensionless potential and length is measured in units of the Debye screening length $d$ (\ref{spectrum}).  These units are employed throughout the rest of the paper;  the electron density is measured in units of $n_{0}$.  The first integral of (\ref{static_Poisson}) is given by
\begin{equation}
\label{static_Sagdeev}
\frac{\phi'^{2}}{2}+\mathcal{S}(\phi)=E,~~~\mathcal{S}(\phi)=\frac{1}{3}+\phi-\frac{1}{3}(1+2\phi)^{3/2}
\end{equation} 
The SP $\mathcal{S}(\phi)$ has a maximum at $\phi=0$, and for the $E=0$ motion the fictitious particle spends most of its time in the vicinity of $\phi=0$ except for rapid excursion to $\phi=-1/2$ followed by rapid return to $\phi=0$.  This is a "dark" soliton, sharp localized minimum of the potential (and density) accompanied by nullification of the electron density at $\phi=-1/2$;  there is a net positive areal charge density associated with the soliton.  Depression of the potential corresponds to a potential energy maximum for the electrons which is consistent with lower than background electron density at the soliton. 

Numerical analysis of the QHD model for the $\gamma=3$ equation of state (\ref{equation_of_state}) also reveals the existence of dark solitons \cite{Shukla} which are qualitatively different from ours.  Electron density was found to have a minimum flanked by two symmetric localized nearby maxima ("shoulders") where the density is higher than that of the background.  Surprisingly, the potential was found to be a localized \textit{positive} maximum for which no classical explanation can be given.  Inspection of the reported soliton profiles \cite{Shukla} shows that as the quantum parameter $\mathcal{A}$ (\ref{quantum}) decreases, the density drop sharpens, its shoulders become less pronounced while the potential spike progressively decreases in magnitude.  Since for $\mathcal{A}=0$ equations of the QHD model reduce to Eq.(\ref{static_Poisson}), we conclude that there exists a critical value of $\mathcal{A}=\mathcal{A}_{c}$ at which the potential changes sign, the soliton undergoes qualitative change crossing over to our result as $\mathcal{A\rightarrow} 0$.  It remains to be seen whether the existence of this transition is the property of the QHD model or valid more generally.                   

We now show that the two analyzed examples are special cases of a general theory whose central parameter is the Mach number $\mathcal{M}=u/s$.  Specifically, $\mathcal{M}=\infty$ describes the $\zeta=const$ situation while finite compressibility static case is the $\mathcal{M}=0$ limit;  the character of plasma oscillations changes qualitatively at $\mathcal{M}=1$. 

Expanding both sides of Eq.(\ref{combined}) into a Taylor series in $\nu=(n-n_{0})/n_{0}$ and keeping the terms not exceeding the second order we obtain
\begin{equation}
\label{quadratic_equation}
\frac{1}{2}\left [3\mathcal{M}^{2}+\frac{n_{0}^{2}}{ms^{2}}\left (\frac{\partial^{2}\zeta}{\partial n^{2}}\right )_{0}\right ]\nu^{2}-(\mathcal{M}^{2}-1)\nu-\phi=0
\end{equation}    
We observe that the coefficient of the linear in $\nu$ term changes sign at $\mathcal{M}=1$ signaling that this value plays a special role.  In its vicinity the coefficient of the $\nu^{2}$ term can be evaluated at $\mathcal{M}=1$ giving a constant that weakly depends on the equation of state.  For the sake of simplicity, we set this coefficient at $1/2$.  Then substituting solutions to (\ref{quadratic_equation}) into the Poisson equation (\ref{1d_Poisson}) we find 
\begin{eqnarray}
\label{Poisson_M=1}
\phi''&=&\mathcal{M}^{2}-1\pm \sqrt{(\mathcal{M}^{2}-1)^{2}+2\phi},\nonumber\\
\phi&\geqslant&-\frac{(\mathcal{M}^{2}-1)^{2}}{2}
\end{eqnarray}   
Since $\phi=0$ is the zero of the right-hand side in (\ref{Poisson_M=1}), the upper sign describes the $\mathcal{M}<1$ regime while the lower corresponds to $\mathcal{M}>1$.  The first integral of Eq.(\ref{Poisson_M=1}) is still given by Eq.(\ref{static_Sagdeev}) except that the SP is now
\begin{equation}
\label{M=1_Sagdeev}
\mathcal{S}(\phi)=(1-\mathcal{M}^{2})\phi+\frac{(1-\mathcal{M}^{2})^{3}\mp [(\mathcal{M}^{2}-1)^{2}+2\phi ]^{3/2}}{3}
\end{equation}
where the sign convention corresponds to that in Eq.(\ref{Poisson_M=1}).  Evolution of the SP (\ref{M=1_Sagdeev}) with the Mach number $\mathcal{M}$ is shown in Figure 2:  
\begin{figure}
\includegraphics[width=1.0\columnwidth, keepaspectratio]{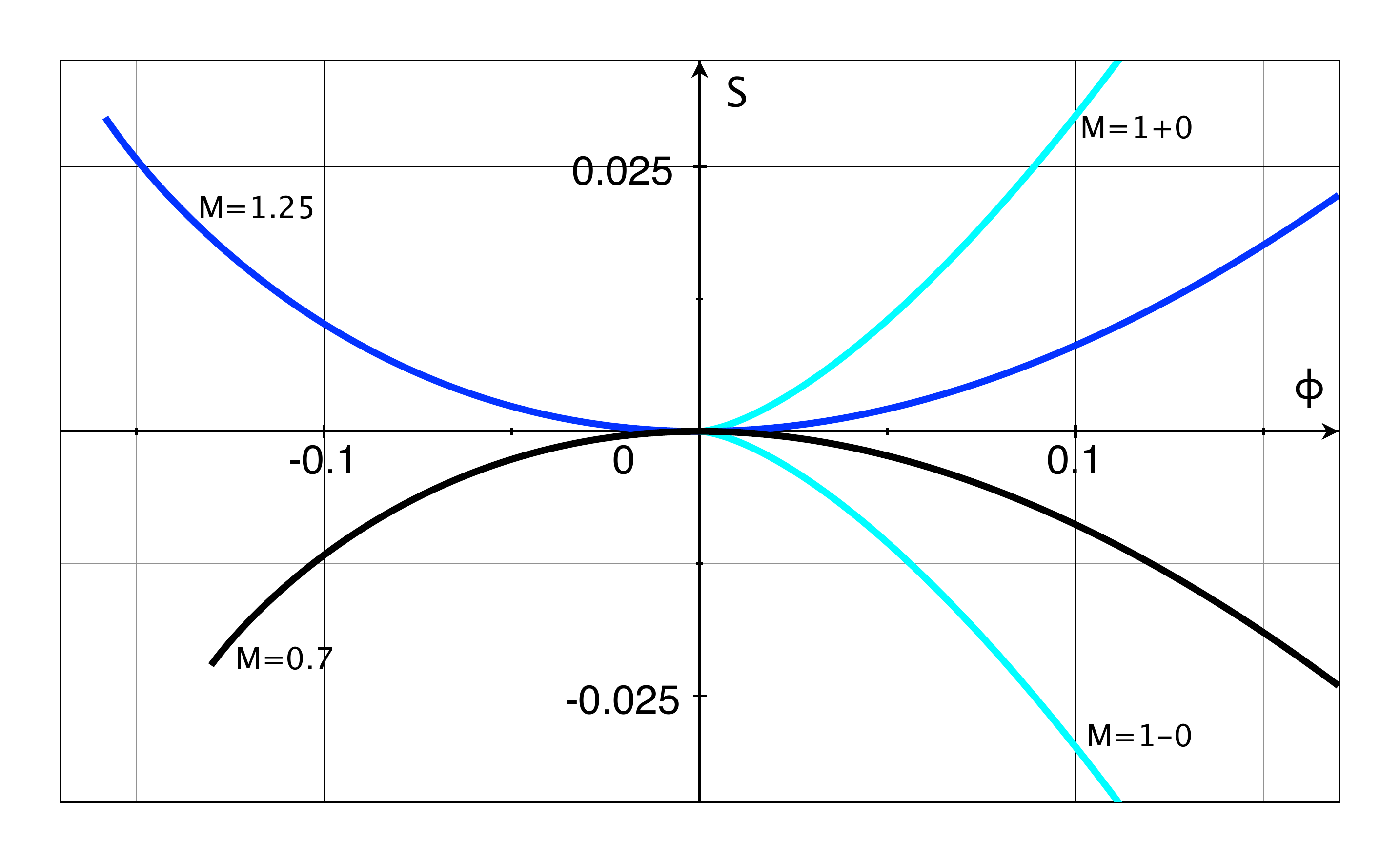} 
\caption{(Color online) Evolution of the Sagdeev potential (\ref{M=1_Sagdeev}) of a generic electron gas with the Mach number $\mathcal{M}$ in the vicinity of $\mathcal{M}=1$.}
\end{figure}  

For $\mathcal{M}<1$ (upper sign in (\ref{M=1_Sagdeev})) the SP has a maximum at $\phi=0$.  The soliton is now a shallow and narrow traveling minimum of the potential of amplitude $(1-\mathcal{M}^{2})^{2}/2$ and width of the order $(1-\mathcal{M}^{2})^{1/2}$.  It is a "grey" soliton as the electron density at its core, $\mathcal{M}^{2}$, is only slightly depressed compared to unity.  

For $\mathcal{M}>1$ (lower sign in (\ref{M=1_Sagdeev})) the SP has a minimum at $\phi=0$ and the phase portrait of the system is qualitatively captured by Figure 1 ($\psi\rightarrow \phi$).  There again exists a marginal trajectory (the counterpart of the blue curve of Figure 1) where the potential reaches its most negative value of $\phi=-(\mathcal{M}^{2}-1)^{2}/2$ for the first time.  However this is no longer an onset of the wave breaking effect because compared to unity the density in the troughs of the potential is only slightly elevated to $\mathcal{M}^{2}$.  This also applies to the counterpart of the dashed trajectory in Figure 1: wave of potential with sharp troughs and smooth crests is now accessible.  This is "deepest" possible wave of the potential.  As the Mach number approaches unity from above, magnitude of the negative potential excursion in the wave tends to zero. For the $\mathcal{M}=1+0$ case (see Figure 2) the wave of the potential is positive everywhere except for the troughs which are zeros of the potential.      

While the analysis relying on the SP (\ref{M=1_Sagdeev}) holds only in the vicinity of $\mathcal{M}=1$ where the wave amplitudes are small, its physics implications do not suffer from this limitation.  This claim can be substantiated by looking at the case of the $\gamma=3$ equation of state (\ref{equation_of_state}) which is explicitly solvable for arbitrary $\mathcal{M}$.  Indeed solving Eq.(\ref{combined}) relative to the density and substituting the outcome into the Poisson equation (\ref{1d_Poisson}) we find
\begin{eqnarray}
\label{M_Poisson}
\phi''&=&\left [\frac{\mathcal{M}^{2}+1}{2}+\phi\pm\sqrt{\left (\frac{\mathcal{M}^{2}+1}{2}+\phi\right )^{2}-\mathcal{M}^{2}}\right ]^{1/2}-1,\nonumber\\
\phi&\geqslant&-\frac{(\mathcal{M}-1)^{2}}{2}
\end{eqnarray}  
where the upper (lower) sign corresponds to $\mathcal{M}<1$ ($\mathcal{M}>1$).  Integrating the right-hand side of (\ref{M_Poisson}) the SP can be recovered.  Since analytic expression for $\mathcal{S}(\phi)$ is cumbersome and not particularly illuminating, we chose to only display its dependence on the Mach number which is shown in Figure 3:  
\begin{figure}
\includegraphics[width=1.0\columnwidth, keepaspectratio]{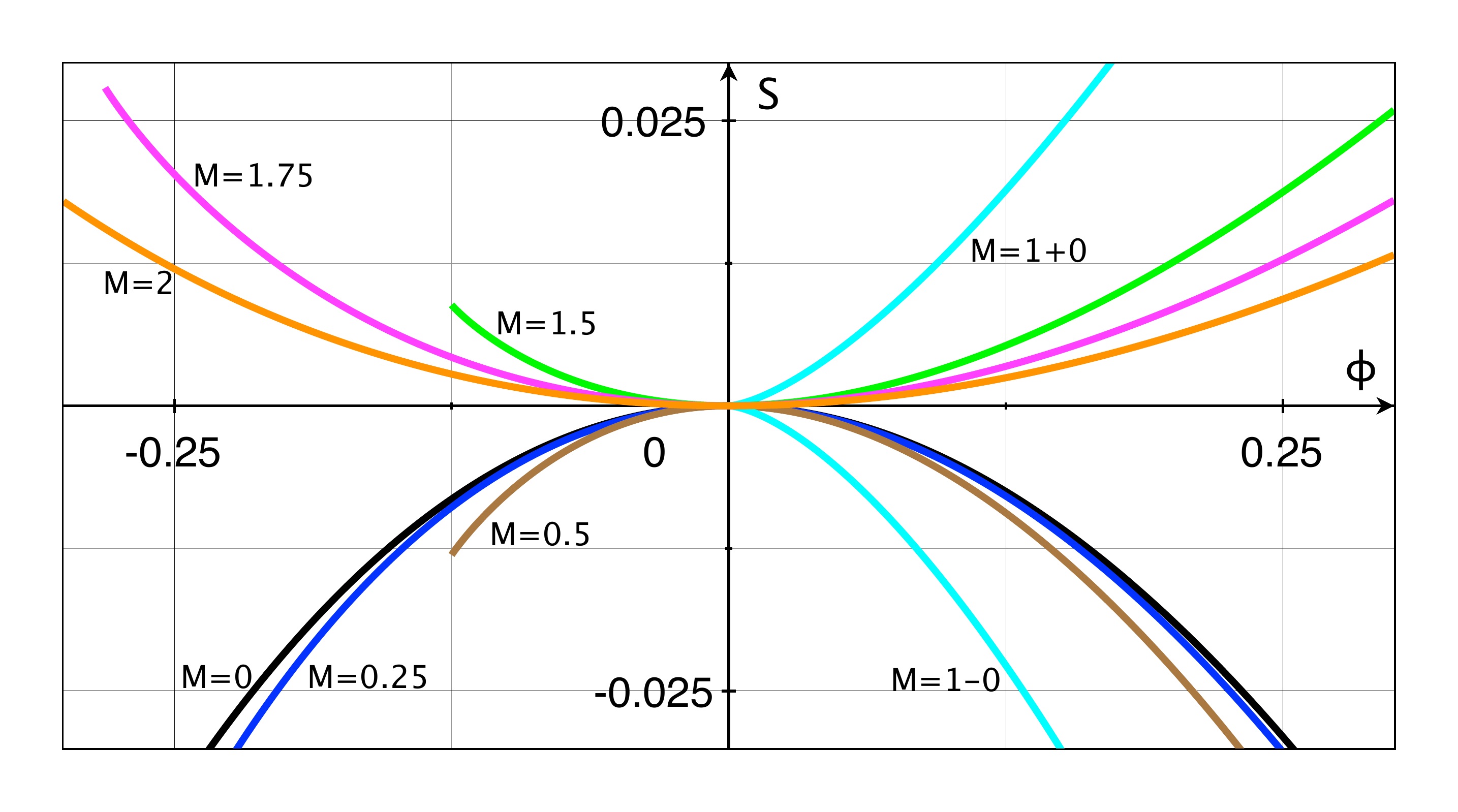} 
\caption{(Color online) Evolution of the Sagdeev potential of the $\gamma=3$ electron gas (\ref{equation_of_state}) with the Mach number $\mathcal{M}$.}
\end{figure}  
 
For $\mathcal{M}<1$ the SP has a maximum at $\phi=0$ which allows soliton solutions.  They are localized minima of the potential and density attaining at soliton's core the values of $-(\mathcal{M}-1)^{2}/2$ and $\sqrt{\mathcal{M}}$, respectively.  At rest, $\mathcal{M}=0$, the soliton is dark while in motion, $\mathcal{M}\neq 0$, it is grey;  solitons cease to exist as $\mathcal{M}\rightarrow 1-0$.  Similar behavior is displayed by solitons of a one-dimensional Bose liquid \cite{KNSQ}. 

For $\mathcal{M}>1$ the SP has a minimum at $\phi=0$ which allows smooth periodic solutions;  in the small amplitude limit they become harmonic.   Additionally periodic solutions with sharp troughs and smooth crests of the potential are allowed with the potential at the troughs given by $-(\mathcal{M}-1)^{2}/2$.  Restoring physical units and taking the $\mathcal{M}\rightarrow \infty$ limit then reproduces the original AL result $\varphi=-mu^{2}/2e$ \cite{AL} which for generic case sets an upper limit on how "deep" the wave of the potential can become.  Even though the electrons accumulate in the troughs with density $\sqrt{\mathcal{M}}$, the wave breaking effect is only present in the $\mathcal{M}\rightarrow\infty$ limit.


\begin{thebibliography}{20}
\bibitem{note}  In an electron (hole) doped semiconductor the right-hand side in (\ref{Poisson}) has to be divided by the dielectric constant of the pristine material which for simplicity is omitted.
\end{thebibliography}
\end{document}